\documentclass[aps,prb,amsmath,amssymb,reprint]{revtex4-1}
\usepackage{graphicx}
\usepackage{bm}

\begin{document}

\title{Nematic and time-reversal breaking superconductivities\\
  coexisting with quadrupole order
  in a $\Gamma_3$ system}

\author{Katsunori Kubo}
\affiliation{Advanced Science Research Center,
  Japan Atomic Energy Agency, Tokai, Ibaraki 319-1195, Japan}

\date{\today}

\begin{abstract}
We discuss superconductivity in a model on a cubic lattice
for a $\Gamma_3$ non-Kramers system.
In previous studies, it is revealed that
$d$-wave superconductivity with $E_g$ symmetry occurs in a wide parameter range
in a $\Gamma_3$ system.
Such anisotropic superconductivity can break the cubic symmetry
of the lattice.
In a $\Gamma_3$ system, the quadrupole degrees of freedom are active
and the effect of the cubic symmetry breaking should be important.
Here, we investigate the coexisting states of the $d$-wave superconductivity
and quadrupole order by a mean-field theory.
In particular, we discuss possible competition and cooperation
between the superconductivity and quadrupole order depending on types of them.
We find nematic superconductivity breaking the cubic symmetry
and coexisting with quadrupole order.
In the present model, we also find $d+id$ superconductivity,
which breaks time-reversal symmetry but retains the cubic symmetry.
We also discuss the effects of uniaxial stress on
these superconducting states.
\end{abstract}

\maketitle

\section{Introduction}
In the vicinity of antiferromagnetically ordered phases,
superconductivity is often found such as
in cuprate high-temperature superconductors~\cite{Pickett1989,Dagotto1994},
in Fe-based superconductors~\cite{Ishida2009,Stewart2011},
and in heavy-fermion materials~\cite{Onuki2004,Lohneysen2007}.
Superconductivity in these materials cannot be explained
by the conventional phonon-mediated pairing mechanism, and they are called 
unconventional superconductors~\cite{Sigrist1991,Norman2011}.
The mechanism of the unconventional superconductivity in these materials
has been a central issue in condensed matter physics.
In particular, the spin-fluctuation-mediated
superconducting mechanism has been widely discussed.

In addition, the orbital degrees of freedom may also play key roles
in the superconductivity of orbitally degenerate systems.
The importance of the orbital degrees of freedom was suggested
for $f$-electron superconductors~\cite{Takimoto2002,Takimoto2003,
  Kubo2006JPSJ,Kubo2007JMMM}.
For the Fe-based superconductors,
the importance of the orbital fluctuations
has also been discussed~\cite{Kontani2010,Yanagi2010}.
Then, the interplay between the spin and orbital degrees of freedom
became an important issue in the field of the unconventional superconductivity.

It is also interesting
to explore possibility of unconventional superconductivity
without spin degrees of freedom.
For this purpose,
an $f$-electron system with the $\Gamma_3$ non-Kramers doublet state
under a cubic crystalline electric field (CEF) is a plausible candidate.
The $\Gamma_3$ state has the same symmetry as the spinless $e_g$ electron
and possesses quadrupole and octupole moments
but does not have the dipole moment.
Thus, the $\Gamma_3$ system can be regarded as an ideal system
to investigate orbital physics
and there may be a route to unconventional superconductivity
other than the spin fluctuation mechanism.

In actual $\Gamma_3$ systems,
superconductivity has been reported
in PrT$_2$X$_{20}$ (T = transition metal element, X = Zn, Al).
In this series of compounds,
the CEF ground state of the $f^2$ electronic configuration in a Pr$^{3+}$ ion
is the $\Gamma_3$ doublet~\cite{Koseki2011,Onimaru2011,Matsushita2011,
  Sakai2011,Ishii2011,Sato2012,Ishii2013,Iwasa2013,Hamamoto2017}
(strictly, in PrRh$_2$Zn$_{20}$,
the CEF ground state is the $\Gamma_{23}$ doublet
due to symmetry lowering at the Pr site
induced by a structural transition~\cite{Iwasa2013}).
In these materials, quadrupole order often realizes
and the relation between the superconductivity
and the quadrupole degrees of freedom has been discussed.
In PrIr$_2$Zn$_{20}$~\cite{Onimaru2010,Onimaru2011,Ishii2011,Iwasa2017}
and PrV$_2$Al$_{20}$~\cite{Sakai2011,Tsujimoto2014},
superconductivity takes place
below the antiferroquadrupole (AFQ) ordering temperature $T_{\text{AFQ}}$.
For PrIr$_2$Zn$_{20}$,
the order parameter of the AFQ ordering
is determined to be $O^2_2=x^2-y^2$~\cite{Iwasa2017}.
On the other hand, superconductivity takes place
below the ferroquadrupole (FQ) ordering temperature $T_{\text{FQ}}$
in PrTi$_2$Al$_{20}$~\cite{Koseki2011,Sakai2011,Sato2012,Ito2011,Sakai2012,
  Matsubayashi2012,Matsubayashi2014,Taniguchi2016}.
The order parameter of the FQ ordering in this compound
is determined to be $O^0_2=3z^2-r^2$~\cite{Sato2012,Taniguchi2016}.
In PrRh$_2$Zn$_{20}$,
superconductivity occurs simultaneously
with AFQ ordering~\cite{Onimaru2012,Ishii2013}.

To develop a theory for the $\Gamma_3$ systems,
we have constructed models for $f$ electrons
with the total angular momentum $j=5/2$.
In these models, we have introduced effective interactions
between $f$ electrons to realize the $\Gamma_3$ CEF state
as the ground state of an $f^2$ ion.
First, we have derived the multipole interactions
in the strong coupling limit~\cite{Kubo2017,Kubo2018JPCS}.
Then, we have found that two-orbital models
are insufficient to discuss multipole physics of the $\Gamma_3$ systems.
On the other hand, the derived multipole interactions for a three-orbital model
depend reasonably on lattice structure.

We have also investigated superconductivity in the $\Gamma_3$ system
by applying a random phase approximation (RPA)
to the three-orbital model~\cite{Kubo2018JPSJ,Kubo2018AIPA}.
Then, we have found instability to the $E_g$ spin-singlet superconductivity.
Such $d$-wave superconductivity is naturally expected in this model
by the following reason.
We have introduced an antiferromagnetic interaction
between the orbitals belonging to $\Gamma_7$ and $\Gamma_8$ symmetry
to stabilize the $f^2$-$\Gamma_3$ CEF state.
This interaction also works as an onsite spin-singlet pairing interaction
between electrons in these orbitals.
The orbital symmetry can be rewritten as
$\Gamma_7=\Gamma_2 \times \Gamma_6$ and
$\Gamma_8=\Gamma_3 \times \Gamma_6$,
where $\Gamma_6$ describes the Kramers or spin degeneracy.
Thus, the interorbital spin-singlet pairing state composed
of the $\Gamma_7$ and $\Gamma_8$ orbitals on the same site has
the $E_g$ $(=\Gamma_3=\Gamma_2 \times \Gamma_3)$ symmetry.
Such anisotropic superconductivity originating from the orbital anisotropy
was already discussed for a simple two-orbital Hubbard model~\cite{Kubo2007PRB}.

When anisotropic superconductivity occurs in a cubic system,
the lattice symmetry lowers with some exceptions
such as a $d_{x^2-y^2}+id_{3z^2-r^2}$ state~\cite{Sigrist1991}.
In this sense, $d_{x^2-y^2}$ and $d_{3z^2-r^2}$ superconductivities
may be called as nematic superconductivity
as in Cu$_x$Bi$_2$Se$_2$~\cite{Fu2014,Matano2016,Pan2016,Yonezawa2017}
and Nb$_x$Bi$_2$Se$_2$~\cite{Asaba2017}.
In a system with the quadrupole degrees of freedom like the $\Gamma_3$ system,
the FQ moment should be finite
in the nematic superconducting state.
Thus, we expect a coexistent state of
the $d$-wave superconductivity and FQ order
in the $\Gamma_3$ system~\cite{Kubo2018AIPA}.
In addition, superconductivity in PrT$_2$X$_{20}$ occurs
in the quadrupole ordered phases in most cases.
Thus, the coexistence with quadrupole order is important
to understand the superconductivity in the $\Gamma_3$ system.

However, superconductivity in the $\Gamma_3$ system
has been explored theoretically only by RPA in the normal state
and characteristics of the coexistent phase is not yet clarified.
Thus, theoretical studies for the ordered phase are highly desired.
In addition to the coexistent phase,
superconducting phase without quadrupole order is also interesting.
The $E_g$ superconducting state is degenerate
and the time reversal breaking superconductivity of $d+id$
is possible as is discussed for
URu$_2$Si$_2$~\cite{Yano2008,Kasahara2007,Kittaka2016},
graphene~\cite{Nandkishore2012,Kiesel2012,Black-Schaffer2014},
and SrPtAs~\cite{Goryo2012}.
If the $d+id$ state realizes,
the superconducting transition temperature $T_{\text{SC}}$
can be increased under uniaxial stress~\cite{Sigrist1991},
as in a $p+ip$ state discussed for Sr$_2$RuO$_2$~\cite{Hicks2014,Steppke2017}.

In this paper, we investigate
the coexistent state of the $d$-wave superconductivity
and quadrupole order in the $\Gamma_3$ system
by applying a mean-field theory to the three orbital model.
In Sec.~\ref{model}, we introduce the mean-field Hamiltonian.
In Sec.~\ref{results}, we show the calculated results
for the coexistence phases of the superconductivity and quadrupole order.
We also consider effects of uniaxial stress
by introducing an external field to the quadrupole moment
in Sec.~\ref{stress}.
We summarize the paper in Sec.~\ref{summary}

\section{Mean-field Hamiltonian}\label{model}
\subsection{Basis states and kinetic energy term}
In this study, we consider the $f$-electron states
with the total angular momentum $j=5/2$ as the one-electron states.
The $j=7/2$ states have higher energy due to the spin-orbit interaction
and we simply ignore them.
The $j=5/2$ states split into the $\Gamma_7$ and $\Gamma_8$ levels
under a cubic CEF.
The $\Gamma_8$ states at site $\bm{r}$ are given by
\begin{align}
  c^{*}_{\bm{r} \alpha \uparrow} |0 \rangle
  &=
  \frac{1}{\sqrt{6}}
  ( \sqrt{5}a^{*}_{\bm{r} 5/2}
  +a^{*}_{\bm{r} -3/2})|0 \rangle,\\
  c^{*}_{\bm{r} \alpha \downarrow} |0 \rangle
  &= 
  \frac{1}{\sqrt{6}}
  ( \sqrt{5}a^{*}_{\bm{r} -5/2}
  +a^{*}_{\bm{r} 3/2} )|0 \rangle,\\
  c^{*}_{\bm{r} \beta \uparrow} |0 \rangle
  &=
  a^{*}_{\bm{r} 1/2}|0 \rangle,\\
  c^{*}_{\bm{r} \beta \downarrow} |0 \rangle
  &=
  a^{*}_{\bm{r} -1/2}|0 \rangle,
\end{align}
where
$a^{*}_{\bm{r} j_z}$ is the creation operator
of the electron with the $z$-component $j_z$ of the total momentum
at site $\bm{r}$
and $|0\rangle$ denotes the vacuum state.
We will use $\dagger$ to denote the Hermitian conjugate of a matrix,
and here, we have used $*$ to represent the creation operators
to avoid confusion.
The $\Gamma_7$ states are given by
\begin{align}
  c^{*}_{\bm{r} \gamma \uparrow} |0 \rangle
  &=
  \frac{1}{\sqrt{6}}
  (a^{*}_{\bm{r} 5/2}
  -\sqrt{5} a^{*}_{\bm{r} -3/2})|0 \rangle,\\
  c^{*}_{\bm{r} \gamma \downarrow} |0 \rangle
  &=
  \frac{1}{\sqrt{6}}
  (a^{*}_{\bm{r} -5/2}
  -\sqrt{5} a^{*}_{\bm{r} 3/2})|0 \rangle.
\end{align}
In these states,
$\sigma = \uparrow$ or $\downarrow$ denotes the Kramers degeneracy
of the one-electron states.
While it is not a real spin due to the spin-orbit coupling,
we call it spin for simplicity in the following.

In general, the conduction bands near the Fermi level are composed
not only of the $f$ orbital.
However, in the present study, we consider an $f$-orbital only model
as one of the simplest models to describe the $\Gamma_3$ systems.
The influence of the other orbitals may be partially included
in the effective $f$-electron hopping~\cite{Kubo2005PRBB}.
We consider the $f$-electron hopping
through $\sigma$ bonding $(ff\sigma)$
on a simple cubic lattice~\cite{Hotta2003,Kubo2005PRB}.
Then, the kinetic energy term is given by
\begin{equation}
  H_{\text{kin}}
  =\sum_{\bm{k} \tau \tau' \sigma}
  c^{*}_{\bm{k} \tau \sigma}
  \xi_{\bm{k} \tau \tau'}
  c_{\bm{k} \tau' \sigma},
\end{equation}
with
\begin{equation}
  \xi_{\bm{k}}
  =
  \begin{pmatrix}
    3t(c_x+c_y) & -\sqrt{3}t(c_x-c_y) & 0 \\
    -\sqrt{3}t(c_x-c_y) & t(c_x+c_y+4c_z) & 0 \\
    0 & 0 & 0
  \end{pmatrix},
\end{equation}
where $c_i = \cos k_i$ ($i=x$, $y$, or $z$),
$t=3(ff\sigma)/14$,
and we have set the lattice constant as unity.
The bandwidth is $W=12t$.

\subsection{Mean-field Hamiltonian for quadrupole ordering}
A one-electron operator and its Fourier transformation
are written as
\begin{align}
  \hat{A}(\bm{r})
  &=
  \sum_{\tau \tau' \sigma}
  \tilde{A}_{\tau \tau'}c^*_{\bm{r} \tau \sigma}c_{\bm{r} \tau' \sigma},\\
  \begin{split}
    \hat{A}(\bm{q})
    &=
    \frac{1}{N}
    \sum_{\bm{r}}e^{-i\bm{q}\cdot\bm{r}}\hat{A}(\bm{r})\\
    &=
    \frac{1}{N}
    \sum_{\tau \tau' \sigma \bm{k}}
    \tilde{A}_{\tau \tau'}c^*_{\bm{k} \tau \sigma}c_{\bm{k}+\bm{q} \tau' \sigma},
  \end{split}
\end{align}
respectively.
$N$ is the number of the lattice sites.
In the $\Gamma_3$ CEF state,
the quadrupole moments of $\Gamma_3$ symmetry are active.
The matrices for the $\Gamma_3$ quadrupole moments are given by
\begin{align}
  \tilde{O}^0_2
  &=
  \frac{1}{\sqrt{42}}
  \begin{pmatrix}
    4 & & \sqrt{5} \\
     & -4 & \\
    \sqrt{5} & &
  \end{pmatrix}, \label{eq:O20}\\
  \tilde{O}^2_2
  &=
  \frac{1}{\sqrt{42}}
  \begin{pmatrix}
     & 4 & \\
    4 & & -\sqrt{5} \\
     & -\sqrt{5} &
  \end{pmatrix}, \label{eq:O22}
\end{align}
where we have normalized them so as to satisfy
$\text{Tr}(\tilde{O}_2^0)^2=\text{Tr}(\tilde{O}_2^2)^2=1$.
Here, $\text{Tr}$ denotes the trace of a matrix.
%
The intersite quadrupole interaction is given by
\begin{equation}
  \begin{split}
    H^{\text{(Q)}}_{\text{I}}
    &=
    \sum_{A B (\bm{r},\bm{r}')}
    J_{AB}(\bm{r}-\bm{r}') \hat{A}(\bm{r}) \hat{B}(\bm{r}')\\
    &=
    \frac{N}{2}
    \sum_{A B \bm{q}}
    J_{AB}(\bm{q}) \hat{A}(-\bm{q}) \hat{B}(\bm{q}),
  \end{split}
\end{equation}
where $(\bm{r},\bm{r}')$ denotes a pair of lattice sites and
\begin{equation}
  J_{AB}(\bm{q})=\sum_{\bm{r}}e^{-i\bm{q}\cdot\bm{r}}J_{AB}(\bm{r}).
\end{equation}
To deal with the effect of a uniaxial stress,
we consider an external field to the quadrupole moments:
\begin{equation}
  H^{\text{(Q)}}_{\text{ext}}
  =
  -\sum_{A \bm{r}}
  H_{A}(\bm{r}) \hat{A}(\bm{r})
  =
  -N
  \sum_{A \bm{q}}
  H_{A}(-\bm{q}) \hat{A}(\bm{q}),
  \label{eq:H20}
\end{equation}
where
\begin{equation}
  H_{A}(\bm{q})
  =
  \frac{1}{N}
  \sum_{\bm{r}}
  e^{-i\bm{q}\cdot\bm{r}}
  H_{A}(\bm{r}).  
\end{equation}
For a uniaxial stress along the $z$ direction,
we vary $H_{20} \equiv H_{O^0_2}(\bm{q}=\bm{0})$.

We apply a mean-field approximation to the quadrupole interaction:
\begin{equation}
  \begin{split}
    & H^{\text{(Q)}}_{\text{I}}+H^{\text{(Q)}}_{\text{ext}}\\
    \simeq & \,
    \frac{N}{2}
    \sum_{A B \bm{q}}
    J_{AB}(\bm{q})
    \Bigl(
    \hat{A}(-\bm{q}) \langle \hat{B}(\bm{q}) \rangle
    +\langle \hat{A}(-\bm{q}) \rangle \hat{B}(\bm{q})\\
    &-\langle \hat{A}(-\bm{q}) \rangle \langle \hat{B}(\bm{q}) \rangle
    \Bigr)
    +H^{\text{(Q)}}_{\text{ext}}\\
    = & \,
    \sum_{\bm{q}  \sigma \bm{k}}
    c^{\dagger}_{\bm{k} \sigma}
    \Delta^{\text{Q}}_{\bm{q}} c_{\bm{k}+\bm{q} \sigma}
    +E^{\text{(Q, I)}}_0\\
    = & \,
    H^{\text{(MF, Q)}}_{\text{I}}
    +E^{\text{(Q, I)}}_0.
  \end{split}
\end{equation}
Here, we have introduced the following notations:
\begin{align}
  c_{\bm{k} \sigma}
  &=
  \begin{pmatrix}
    c_{\bm{k} \alpha \sigma} \\ c_{\bm{k} \beta \sigma} \\ c_{\bm{k} \gamma \sigma}
  \end{pmatrix},\\
  \Delta^{\text{Q}}_{\bm{q}}
  &=
  \sum_{B}
  \left[
  \sum_{A}
  J_{AB}(\bm{q})
  \langle \hat{A}(-\bm{q}) \rangle
  -H_{B}(-\bm{q})
  \right]  
  \tilde{B},\\
  E^{\text{(Q, I)}}_0
  &=
  -\frac{N}{2}
  \sum_{A B \bm{q}}
  J_{AB}(\bm{q})
  \langle \hat{A}(-\bm{q}) \rangle \langle \hat{B}(\bm{q}) \rangle.
\end{align}
$\langle \cdots \rangle$ denotes the expectation value.
In the following, we denotes the expectation value
also as $A(\bm{q})=\langle \hat{A}(\bm{q}) \rangle$.
In this study,
we consider ordering with $\bm{q}=0$ and $\bm{q}=\bm{Q}=(\pi,\pi,\pi)$.
Then, the mean-field Hamiltonian is written as
\begin{equation}
  \begin{split}
    & H^{\text{(MF, Q)}}\\
    = & \,
    H_{\text{kin}}+H^{\text{(MF, Q)}}_{\text{I}}\\
    = &
    \sum_{\sigma \bm{k} \in \text{FBZ}}
    (c^{\dagger}_{\bm{k} \sigma} c^{\dagger}_{\bm{k}+\bm{Q} \sigma})
    \begin{pmatrix}
      \xi_{\bm{k}}+\Delta^{\text{Q}}_{\bm{0}} &
      \Delta^{\text{Q}}_{\bm{Q}}\\
      \Delta^{\text{Q}}_{\bm{Q}} &
      \xi_{\bm{k}+\bm{Q}}+\Delta^{\text{Q}}_{\bm{0}}
    \end{pmatrix}
    \begin{pmatrix}
      c_{\bm{k} \sigma} \\
      c_{\bm{k}+\bm{Q} \sigma}
    \end{pmatrix}\\
    = &
    \sum_{\sigma \bm{k} \in \text{FBZ}}
    c^{\dagger}_{\sigma}(\bm{k})
    \xi(\bm{k})
    c_{\sigma}(\bm{k}).
  \end{split}
\end{equation}
The $\bm{k}$ summation runs over the folded Brillouin-zone (FBZ)
of the staggered ordering with $\bm{Q}$.

We also consider onsite Coulomb interactions:
intraorbital Coulomb interactions $U_7$ for the $\Gamma_7$ orbital
and $U_8$ for the $\Gamma_8$ orbitals
and interorbital Coulomb interaction $U'_8$ for the $\Gamma_8$ orbitals.
These interactions are also included within the mean-field approximation.
For this purpose, we define charge operators by the following matrices:
\begin{equation}
  \tilde{n}_{\alpha}=
  \begin{pmatrix}
    1 & & \\
    & 0 & \\
    & & 0
  \end{pmatrix},
  \tilde{n}_{\beta}=
  \begin{pmatrix}
    0 & & \\
    & 1 & \\
    & & 0
  \end{pmatrix},
  \tilde{n}_{\gamma}=
  \begin{pmatrix}
    0 & & \\
    & 0 & \\
    & & 1
  \end{pmatrix}.
\end{equation} 
The mean fields from the Coulomb interactions are written as
the mean fields for charge interactions with
$J_{n_{\alpha} n_{\alpha}}(\bm{q})=J_{n_{\beta} n_{\beta}}(\bm{q})=U_8/2$,
$J_{n_{\gamma} n_{\gamma}}(\bm{q})=U_7/2$, and
$J_{n_{\alpha} n_{\beta}}(\bm{q})=J_{n_{\beta} n_{\alpha}}(\bm{q})=U'_8$.
Then, the effects of the Coulomb interactions can be included
in the above formulation for the quadrupole interaction.
Without the Coulomb interactions,
we find a tendency toward phase separation
in particular for a strong quadrupole interaction.
Since the phase separation is not our concern in this study,
we have introduced the Coulomb interactions to alleviate this tendency.

In actual calculations,
we consider ordering states
with the principal axis along the $z$ direction.
Then, we assume interactions purely for $O_2^0$ or $O_2^2$.
We denote them as
$J_{O_2^0 O_2^0}(\bm{q}) = J_{20}(\bm{q})$
and
$J_{O_2^2 O_2^2}(\bm{q}) = J_{22}(\bm{q})$.

\subsection{Mean-field Hamiltonian for superconductivity}
For superconductivity, we assume the following interaction
for spin-singlet pairing:
\begin{equation}
  \begin{split}
    H^{\text{(SC)}}_{\text{I}}
    =&
    -\frac{1}{N}\sum_{\tau \tau' \bm{k} \bm{k}' \bm{q}}
    \frac{V_{\tau \tau'}(\bm{q})}{4}\\
    &\times (c^{*}_{-\bm{k} \tau \downarrow} c^{*}_{\bm{k}+\bm{q} \tau' \uparrow}
    -c^{*}_{-\bm{k} \tau \uparrow} c^{*}_{\bm{k}+\bm{q} \tau' \downarrow})\\
    &\times
    (c_{\bm{k}'+\bm{q} \tau \uparrow} c_{-\bm{k}' \tau' \downarrow}
    -c_{\bm{k}'+\bm{q} \tau \downarrow} c_{-\bm{k}' \tau' \uparrow}
    )\\
    \simeq &
    -\frac{1}{N}\sum_{\tau \tau' \bm{k} \bm{k}' \bm{q}}
    \frac{V_{\tau \tau'}(\bm{q})}{4}\\
    & \times
    \Bigl[
      (c^{*}_{-\bm{k} \tau \downarrow} c^{*}_{\bm{k}+\bm{q} \tau' \uparrow}
      -c^{*}_{-\bm{k} \tau \uparrow} c^{*}_{\bm{k}+\bm{q} \tau' \downarrow})\\
      & \times
      \langle
      c_{\bm{k}'+\bm{q} \tau \uparrow} c_{-\bm{k}' \tau' \downarrow}
      -c_{\bm{k}'+\bm{q} \tau \downarrow} c_{-\bm{k}' \tau' \uparrow}
      \rangle
      \\
      &+
      \langle
      c^{*}_{-\bm{k} \tau \downarrow} c^{*}_{\bm{k}+\bm{q} \tau' \uparrow}
      -c^{*}_{-\bm{k} \tau \uparrow} c^{*}_{\bm{k}+\bm{q} \tau' \downarrow}
      \rangle\\
      &\times
      (c_{\bm{k}'+\bm{q} \tau \uparrow} c_{-\bm{k}' \tau' \downarrow}
      -c_{\bm{k}'+\bm{q} \tau \downarrow} c_{-\bm{k}' \tau' \uparrow}
      )\\
      &-
      \langle
      c^{*}_{-\bm{k} \tau \downarrow} c^{*}_{\bm{k}+\bm{q} \tau' \uparrow}
      -c^{*}_{-\bm{k} \tau \uparrow} c^{*}_{\bm{k}+\bm{q} \tau' \downarrow}
      \rangle\\
      &\times
      \langle
      c_{\bm{k}'+\bm{q} \tau \uparrow} c_{-\bm{k}' \tau' \downarrow}
      -c_{\bm{k}'+\bm{q} \tau \downarrow} c_{-\bm{k}' \tau' \uparrow}
      \rangle
      \Bigr].
  \end{split}
\end{equation}
The superconducting pairing interaction is symmetric
with respect to the orbital index: $V_{\tau \tau'}(\bm{q})=V_{\tau' \tau}(\bm{q})$.
While we will consider only pairing states with total momentum $\bm{q}=\bm{0}$
in actual calculation,
here we write down equations keeping both pairing states
with $\bm{q}=\bm{0}$ and $\bm{Q}$.
Then, the superconducting pairing interaction term is approximated as
\begin{equation}
  \begin{split}
    H^{\text{(SC)}}_{\text{I}}
    \simeq&
    \sum_{\bm{k} \in \text{FBZ}}
    \left[
      c^{\dagger}_{\uparrow}(\bm{k}) \Delta c^{*}_{\downarrow}(-\bm{k})
      +c^{\text{T}}_{\downarrow}(-\bm{k}) \Delta^{\dagger}c_{\uparrow}(\bm{k})
      \right]\\
    &+E^{\text{(SC, I)}}_{0},
  \end{split}
\end{equation}
where $\text{T}$ denotes the transpose of a matrix,
\begin{equation}
  \Delta
  =
  \begin{pmatrix}
    \Delta^{\text{SC}}_{\bm{0}} & \Delta^{\text{SC}}_{\bm{Q}} \\
    \Delta^{\text{SC}}_{\bm{Q}} & \Delta^{\text{SC}}_{\bm{0}}
  \end{pmatrix},
\end{equation}
\begin{equation}
  \Delta^{\text{SC}}_{\bm{q} \tau \tau'}
  =
  \frac{V_{\tau \tau'}(\bm{q})}{N}\sum_{\bm{k}}
  \frac{1}{2}
  \langle
  c_{\bm{k}+\bm{q} \tau \uparrow} c_{-\bm{k} \tau' \downarrow}
  -c_{\bm{k}+\bm{q} \tau \downarrow} c_{-\bm{k} \tau' \uparrow}
  \rangle,
  \label{eq:Delta}
\end{equation}
and
\begin{equation}
  E^{\text{(SC, I)}}_{0}
  =
  N \sum_{\bm{q} \, \tau \tau', V_{\tau \tau'}(\bm{q}) \ne 0}
  |\Delta^{\text{SC}}_{\tau \tau' \bm{q}}|^2/V_{\tau \tau'}(\bm{q}).
\end{equation}
We call the quantity
without the factor $V_{\tau \tau'}(\bm{q})$ in Eq.~\eqref{eq:Delta}
the pair amplitude.

The total Hamiltonian is approximated as
\begin{equation}
  \begin{split}
    H
    &=
    H_{\text{kin}}+H^{\text{(Q)}}_{\text{I}}+H^{\text{(Q)}}_{\text{ext}}
    +H^{\text{(SC)}}_{\text{I}}\\
    &\simeq
    H^{\text{(MF)}}+\sum_{\bm{k} \in \text{FBZ}}\text{Tr} \xi(\bm{k})
    +E^{\text{(Q, I)}}_{0}+E^{\text{(SC, I)}}_{0}\\
    &=
    H^{\text{(MF)}}+E_0,
  \end{split}
  \label{approxH}
\end{equation}
where the mean-field Hamiltonian is given by
\begin{equation}
  \begin{split}
    & H^{\text{(MF)}} \\
    =&\sum_{\bm{k} \in \text{FBZ}}
    (c^{\dagger}_{\uparrow}(\bm{k}) \
    c^{\text{T}}_{\downarrow}(-\bm{k}) )
    \begin{pmatrix}
      \xi(\bm{k}) & \Delta \\
      \Delta^{\dagger} & -\xi^{\text{T}}(-\bm{k})
    \end{pmatrix}
    \begin{pmatrix}
      c_{\uparrow}(\bm{k}) \\
      c^{*}_{\downarrow}(-\bm{k})
    \end{pmatrix}.
  \end{split}
\end{equation}
We solve the mean-field Hamiltonian self-consistently.
When we obtain different solutions,
we select the state with the lowest free-energy.
For the evaluation of the free-energy,
we need to calculate the constant term $E_0$ in Eq.~\eqref{approxH}.

In previous studies~\cite{Kubo2017,Kubo2018JPCS,Kubo2018JPSJ,Kubo2018AIPA},
we have considered an antiferromagnetic interaction
$H_{J}=J \sum_{\bm{r}} \bm{s}_{\bm{r}7} \cdot \bm{s}_{\bm{r}8}$
between electrons
in the $\Gamma_7$ and $\Gamma_8$ orbitals
to realize the $\Gamma_3$ state as the ground state of an $f^2$ ion,
where
$\bm{s}_{\bm{r} 7}=(1/2) \sum_{\sigma \sigma'}
c^{\dagger}_{\bm{r} \gamma \sigma}
\bm{\sigma}_{\sigma \sigma'} c_{\bm{r} \gamma \sigma'}$
and
$\bm{s}_{\bm{r} 8}=(1/2) \sum_{\sigma \sigma' \nu=\alpha,\beta}
c^{\dagger}_{\bm{r} \nu \sigma} \bm{\sigma}_{\sigma \sigma'} c_{\bm{r} \nu \sigma'}$.
$\bm{\sigma}$ are the Pauli matrices.
This interaction results in $\bm{q}$-independent pairing interactions
$V_{\alpha \gamma}(\bm{q})=V_{\beta \gamma}(\bm{q})=3J/4$.
However, in this study,
we consider ordinary pairing states with total momentum $\bm{q}=\bm{0}$,
and then, we include only
$V_{\alpha \gamma}(\bm{0})$ and $V_{\beta \gamma}(\bm{0})$ for simplicity.
We change these pairing interactions independently
to investigate the relation between the quadrupole ordering
and superconducting symmetry.

The superconducting order parameter $\Delta^{\text{SC}}_{\bm{0} \alpha \gamma}$
transforms as $k_x^2-k_y^2$ under symmetry operations.
We consider this $d_{x^2-y^2}$ superconducting pairing state
and denote $\Delta^{\text{SC}}_{\bm{0} \alpha \gamma}=\Delta_{x^2-y^2}$
and $V_{\alpha \gamma}(\bm{0})=V_{x^2-y^2}$.
Similarly, we consider the $d_{3z^2-r^2}$ superconducting pairing state
with $\Delta^{\text{SC}}_{\bm{0} \beta \gamma} \ne 0$
and denote $\Delta^{\text{SC}}_{\bm{0} \beta \gamma}=\Delta_{3z^2-r^2}$
and $V_{\beta \gamma}(\bm{0})=V_{3z^2-r^2}$.

\section{Quadrupole order and superconductivity}\label{results}
In the following calculations,
we set $U_7=5t$, $U_8=10t$, and $U'_8=5t$.
We vary $V_{x^2-y^2}$, $V_{3z^2-r^2}$, $J_{20}(\bm{q})$, and $J_{20}(\bm{q})$
with $\bm{q}=\bm{0}$ and $\bm{Q}$,
but they are set to zero unless finite values are explicitly given.
To realize the $\Gamma_3$ state as the CEF ground state of an $f^2$ ion,
the following conditions should be satisfied~\cite{Kubo2017}:
$n_7 =
\sum_{\sigma} \langle c^*_{\bm{k} \gamma \sigma} c_{\bm{k} \gamma \sigma} \rangle / N
\simeq 1$
and
$n_8 =
\sum_{\sigma, \tau=\alpha,\beta} \langle c^*_{\bm{k} \tau \sigma} c_{\bm{k} \tau \sigma}
\rangle / N
\simeq 1$.
We tune the chemical potential for each CEF level independently
so as to satisfy the conditions $n_7 = 1$ and $n_8 = 1$.
The lattice size is $N = L \times L \times L$ with $L=12$.
We have found that the finite size effect is weak.
In evaluating transition temperatures,
we extrapolate the order parameters to zero
by using data with temperature intervals of $dT=0.02t$.
The differences in the transition temperatures obtained with $L=8$ and $L=12$
are smaller than the errors in this extrapolation.

\subsection{Temperature dependence of order parameters}
First, we show the temperature dependence of the quadrupole moment
without superconducting interactions.
In an ordered state of $O_2^0$ or $O_2^2$ with any periodicity,
the $z$ direction becomes inequivalent to the $x$ and $y$ directions
and the FQ moment $O_2^0(\bm{0})$
should be induced~\cite{Kubo2018AIPA}.
In particular, $O_2^0(\bm{0})$ is induced by
the FQ order of $O_2^0$ itself,
and as a result,
the transition becomes of first order.
This is also understood from the existence of the third-order term
in the Landau free-energy~\cite{Hattori2014,Lee2018}.
Due to the self-induced nature of the $O_2^0(\bm{0})$ order,
the transition temperature of $O_2^0(\bm{0})$ is higher
than that of $O_2^2(\bm{0})$
as long as $J_{20}(\bm{0})=J_{22}(\bm{0})$,
under which the cubic symmetry
of the interaction term is retained~\cite{Sakai2003}.
Indeed, the order parameter of the FQ order in PrTi$_2$Al$_{20}$
is determined to be $O_2^0$~\cite{Sato2012,Taniguchi2016}.
However, in the following, we change $J_{20}(\bm{0})$ and $J_{22}(\bm{0})$
independently to clarify the relation between superconductivity
and kinds of the quadrupole order.

\begin{figure}
  \includegraphics[width=0.99\linewidth]
      {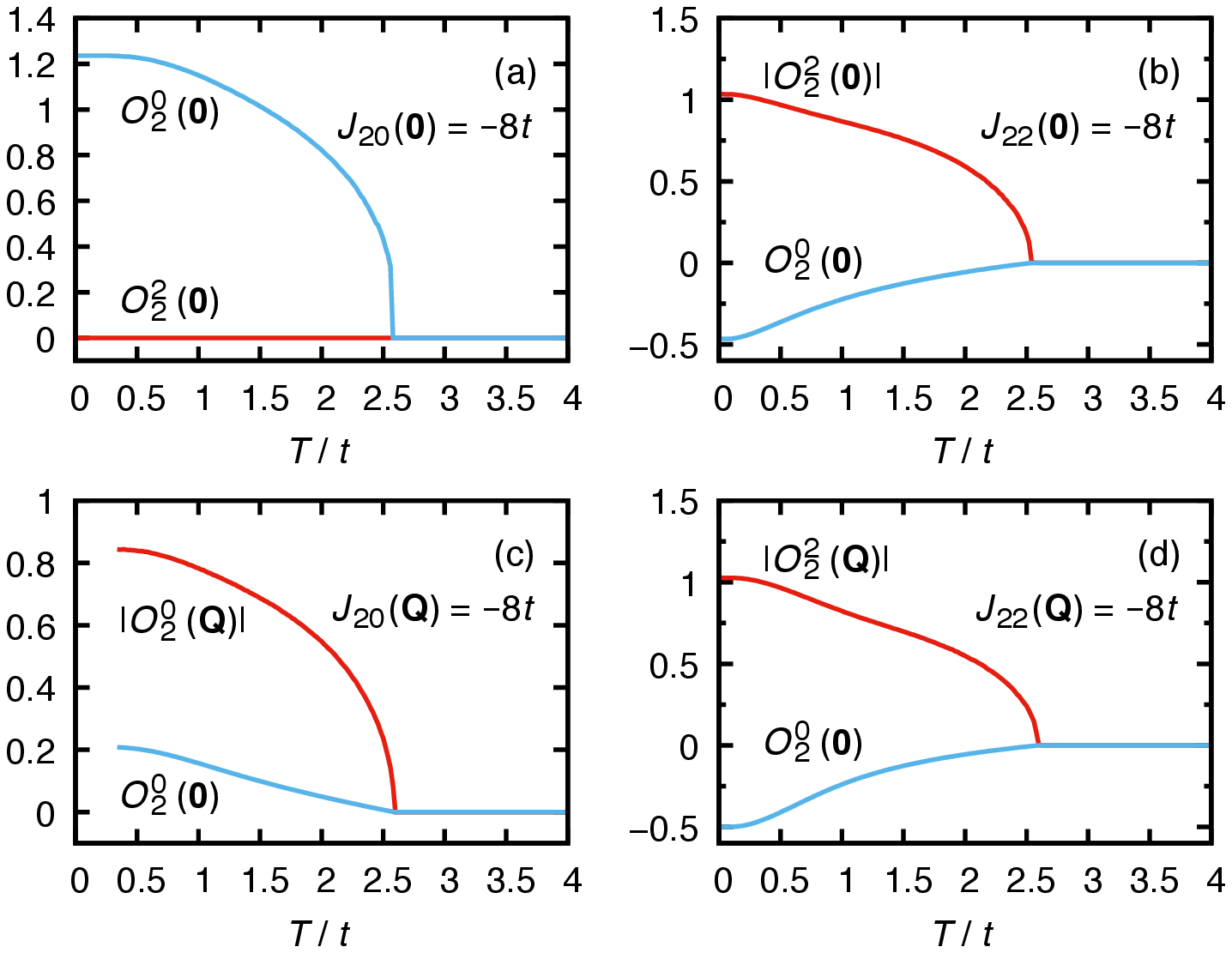}%
  \caption{
    Temperature dependence of the quadrupole order parameters
    (a) for $J_{20}(\bm{0})=-8t$,
    (b) for $J_{22}(\bm{0})=-8t$,
    (c) for $J_{20}(\bm{Q})=-8t$,
    and
    (d) for $J_{22}(\bm{Q})=-8t$.
  \label{quadrupole_order}}
\end{figure}

In Fig.~\ref{quadrupole_order}(a),
we show temperature dependence of $O_2^0(\bm{0})$
for $J_{20}(\bm{0})=-8t$.
As discussed above, the transition is of first order
and $O_2^0(\bm{0})$ jumps at the transition temperature.
For comparison, we also show $O_2^2(\bm{0})$ but it remains zero.
Note that the solutions with positive and negative $O_2^0$ are not equivalent.
In the present model, we find that the solution with positive $O_2^0$
has lower free-energy.
In Fig.~\ref{quadrupole_order}(b),
we show temperature dependence of the order parameter $O_2^2(\bm{0})$
for $J_{22}(\bm{0})=-8t$.
The transition temperature is slightly lower than
that in Fig.~\ref{quadrupole_order}(a).
As discussed above, $O_2^0(\bm{0})$ is induced below the transition temperature.
The induced $O_2^0(\bm{0})$ is negative.
As for AFQ ordering cases,
we show results for $J_{20}(\bm{Q})=-8t$ and for $J_{22}(\bm{Q})=-8t$
in Fig.~\ref{quadrupole_order}(c) and (d), respectively.
The induced moment $O_2^0(\bm{0})$ has the opposite sign
between these two cases.

\begin{figure}
  \includegraphics[width=0.99\linewidth]
      {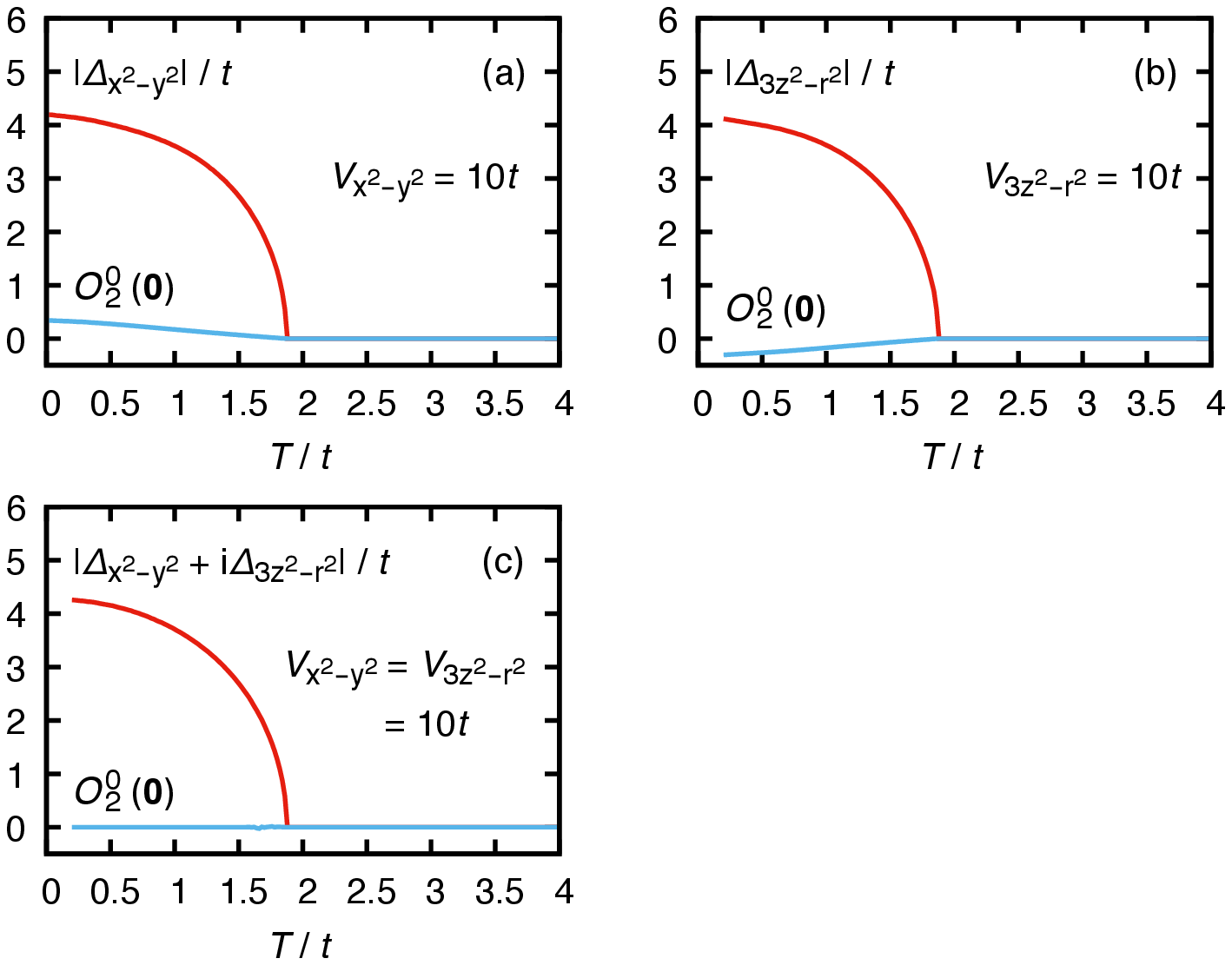}%
  \caption{
    Temperature dependence of the superconducting order parameters
    (a) for $V_{x^2-y^2}  = 10t$,
    (b) for $V_{3z^2-r^2} = 10t$,
    and
    (c) for $V_{x^2-y^2}  = V_{3z^2-r^2} = 10t$.
    The induced $O_2^0(\bm{0})$ is also shown.
  \label{SC_order}}
\end{figure}

Next, we show the temperature dependence
of the superconducting order parameters
without quadrupole interactions.
Although we should set $V_{x^2-y^2}=V_{3z^2-r^2}$ to retain the cubic symmetry
in the interaction term,
we change them independently to understand the relation
between superconducting symmetry and quadrupole order.
Figure~\ref{SC_order}(a) shows temperature dependence
of $\Delta_{x^2-y^2}$ for $V_{x^2-y^2}=10t$.
Below the transition temperature, positive $O_2^0(\bm{0})$ is induced.
For $V_{3z^2-r^2}=10t$ [Fig.~\ref{SC_order}(b)],
$d_{3z^2-r^2}$ superconductivity occurs
at the same transition temperature as in Fig.~\ref{SC_order}(a).
The induced $O_2^0(\bm{0})$ has the opposite sign
between the $d_{x^2-y^2}$ and $d_{3z^2-r^2}$ superconducting states.
We can understand this from the Ginzburg--Landau theory.
At least around the transition temperature,
i.e., when the order parameters are sufficiently small,
the coupling constant of the superconducting order parameter to $O_2^0(\bm{0})$
in the Ginzburg--Landau free-energy has the opposite sign between
the $d_{x^2-y^2}$ and $d_{3z^2-r^2}$ superconducting states~\cite{Sigrist1991}.

Note also that the $d_{3z^2-r^2}$ superconducting state accompanying
the FQ order of $O_2^0$ mixes with $s$-wave superconductivity
since $3z^2-r^2$ is an identity representation in the tetragonal phase.
When we consider the $s$-wave superconducting interaction
and if the FQ moment of $O_2^0$ develops sufficiently,
the effect of the hybridization between these superconducting states
may become strong.
For example, line nodes in the gap function
in the $d_{3z^2-r^2}$ superconductivity can disappear at lower temperatures
due to this hybridization.

For $V_{x^2-y^2}=V_{3z^2-r^2}$,
according to the Ginzburg--Landau theory~\cite{Sigrist1991},
superconducting symmetry is $d_{x^2-y^2}$, $d_{3z^2-r^2}$, or
$d_{x^2-y^2}+id_{3z^2-r^2}$ depending on the microscopic model.
In the present model,
we find $d_{x^2-y^2}+id_{3z^2-r^2}$ superconductivity
as shown in Fig.~\ref{SC_order}(c).
In this superconducting state,
the cubic symmetry is preserved,
since $d_{x^2-y^2}+id_{3z^2-r^2}
= e^{-2\pi i/3}(d_{y^2-z^2}+id_{3x^2-r^2})
= e^{2\pi i/3}(d_{z^2-x^2}+id_{3y^2-r^2})$.
Then, the quadrupole moment $O_2^0(\bm{0})$ is not induced in this case.
In the $d+id$ superconducting state,
the superconducting gap has only point nodes at $|k_x|=|k_y|=|k_z|$
while line nodes appear
in the $d_{x^2-y^2}$ and $d_{3z^2-r^2}$ superconducting states.
It may be natural that
the superconductivity with a larger gapped portion of the Fermi surface is
more stable at least without quadrupole interactions.

\subsection{Phase diagrams}
In this subsection,
we construct phase diagrams
by changing the strength of the quadrupole interaction and temperature
with fixing the values of the superconducting pairing interactions.
In the following, while we do not explicitly denote,
all the ordered phases accompany a FQ moment of $O_2^0$
except for the pure $d+id$ superconducting phase.

\begin{figure}
  \includegraphics[width=0.99\linewidth]
      {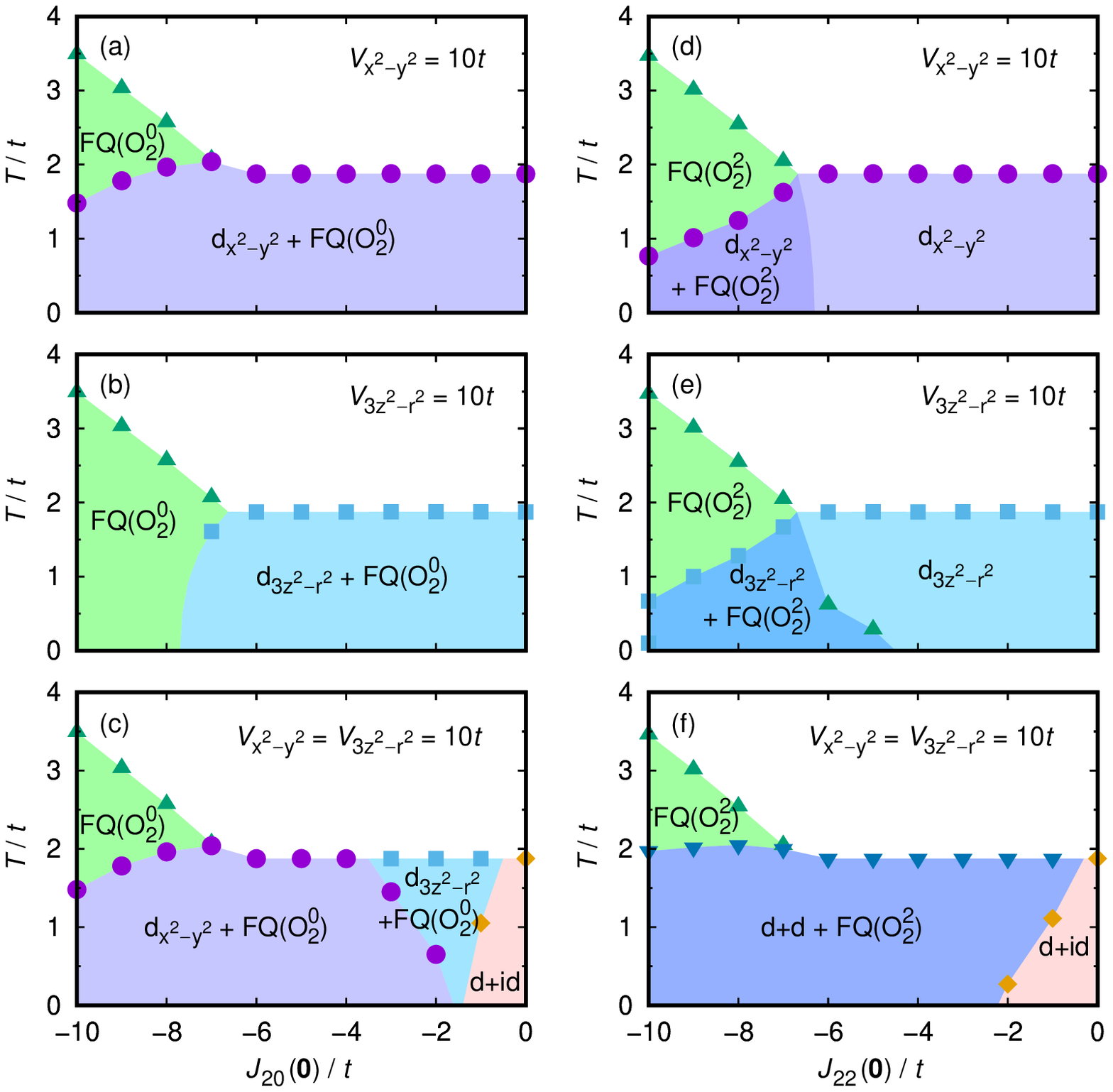}%
  \caption{
    Phase diagrams of superconductivity
    coexisting with FQ order:
    in the $J_{20}(\bm{0})$-$T$ plane
    (a) for $V_{x^2-y^2}  = 10t$,
    (b) for $V_{3z^2-r^2} = 10t$,
    and
    (c) for $V_{x^2-y^2}  = V_{3z^2-r^2} = 10t$;
    in the $J_{22}(\bm{0})$-$T$ plane
    (d) for $V_{x^2-y^2}  = 10t$,
    (e) for $V_{3z^2-r^2} = 10t$,
    and
    (f) for $V_{x^2-y^2}  = V_{3z^2-r^2} = 10t$.
  \label{SC_FQ_order}}
\end{figure}

In Fig.~\ref{SC_FQ_order}(a) and (b),
we show phase diagrams in the $J_{20}(\bm{0})$-$T$ plane
for $V_{x^2-y^2}  = 10t$ and for $V_{3z^2-r^2} = 10t$, respectively.
The region of the $d_{x^2-y^2}$ superconducting phase
in Fig.~\ref{SC_FQ_order}(a) is wider than that of
the $d_{3z^2-r^2}$ superconducting phase in Fig.~\ref{SC_FQ_order}(b).
As shown in Fig.~\ref{quadrupole_order}(a) and  Fig.~\ref{SC_order}(a),
both the pure FQ order of $O_2^0$
and the $d_{x^2-y^2}$ superconducting state accompany positive $O_2^0(\bm{0})$
in the present model and they are expected to be cooperative.
Indeed, the superconducting transition temperature
around $J_{20}(\bm{0})=-7t$ is higher than
that without the quadrupole interaction,
while a too large $|J_{20}(\bm{0})|$ suppresses $T_{\text{SC}}$.
On the other hand, the $d_{3z^2-r^2}$ superconducting state
induces negative $O_2^0(\bm{0})$
and the FQ order of $O_2^0$ is destructive for this superconducting state
as shown in Fig.~\ref{SC_FQ_order}(b).
For $V_{x^2-y^2}  = V_{3z^2-r^2} = 10t$ shown in Fig.~\ref{SC_FQ_order}(c),
the $d+id$ state appears for a small $|J_{20}(\bm{0})|$.
The $d_{x^2-y^2}$ state appears for larger $|J_{20}(\bm{0})|$ in a wide range
since this superconducting state is cooperative with the FQ order of $O_2^0$.
The $d_{3z^2-r^2}$ state appears in a narrow region
between the above two superconducting phases.
In each of the phase diagrams Fig.~\ref{SC_FQ_order}(a) and (b),
only one superconducting phase appears
since the FQ moment of $O_2^0$ is always finite
even without the FQ interaction.
This is in sharp contrast to the other quadrupole interaction cases
we will discuss below.

In PrTi$_2$Al$_{20}$
under hydrostatic pressure~\cite{Matsubayashi2012,Matsubayashi2014},
$T_{\text{SC}}$ seems to have a peak at around pressure
where $T_{\text{SC}}$ reaches $T_{\text{FQ}}$ of $O_2^0$.
The phase diagram Fig.~\ref{SC_FQ_order}(a) or (c)
may be relevant to PrTi$_2$Al$_{20}$
if a hydrostatic pressure mainly affects the intersite quadrupole interactions.
This is a natural assumption
when the superconducting pair is
mainly composed of the same site~\cite{Kubo2018JPSJ}
and insensitive to the change in the distance between lattice sites.

In Fig.~\ref{SC_FQ_order}(d) and (e),
we show phase diagrams in the $J_{22}(\bm{0})$-$T$ plane
for $V_{x^2-y^2}  = 10t$ and for $V_{3z^2-r^2} = 10t$, respectively.
For $V_{3z^2-r^2} = 10t$ at $J_{22}(\bm{0})=-10t$,
superconductivity disappears at a very low temperature.
There are two superconducting phases with and without FQ order of $O_2^2$
in each phase diagram.
The region of the coexisting phase of the $d_{3z^2-r^2}$ superconductivity
with FQ order of $O_2^2$ seems slightly larger than
that of the $d_{x^2-y^2}$ superconductivity.
It may be understood from the fact that
the induced $O_2^0(\bm{0})$
in the $d_{3z^2-r^2}$ superconductivity and FQ order of $O_2^2$ has the same sign.
In the FQ ordered phase of $O_2^2$,
the symmetry is lower than tetragonal, i.e., orthorhombic.
Then, in the coexisting phase of the superconductivity
and the FQ order of $O_2^2$,
both the pair amplitudes for $d_{x^2-y^2}$ and $d_{3z^2-r^2}$ become finite.
Thus, for $V_{x^2-y^2}  = V_{3z^2-r^2} = 10t$ [Fig.~\ref{SC_FQ_order}(f)],
by combining both superconducting order parameters,
$d+d$ superconducting phase realizes in a wide region.

\begin{figure}
  \includegraphics[width=0.99\linewidth]
      {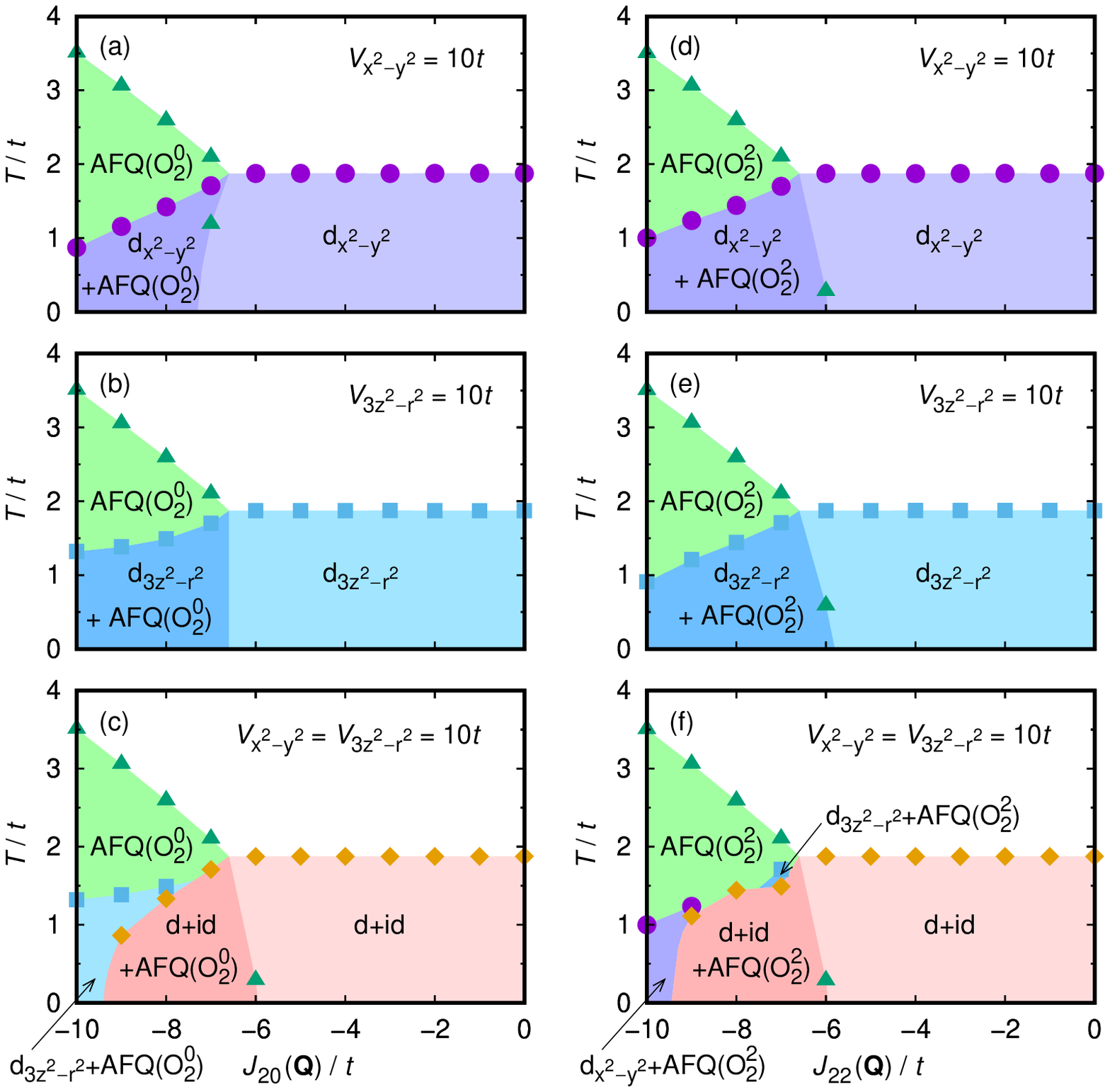}%
  \caption{
    Phase diagrams of superconductivity
    coexisting with AFQ order:
    in the $J_{20}(\bm{Q})$-$T$ plane
    (a) for $V_{x^2-y^2}  = 10t$,
    (b) for $V_{3z^2-r^2} = 10t$,
    and
    (c) for $V_{x^2-y^2}  = V_{3z^2-r^2} = 10t$:
    in the $J_{22}(\bm{Q})$-$T$ plane
    (d) for $V_{x^2-y^2}  = 10t$,
    (e) for $V_{3z^2-r^2} = 10t$,
    and
    (f) for $V_{x^2-y^2}  = V_{3z^2-r^2} = 10t$.
  \label{SC_AFQ_order}}
\end{figure}

For $O_2^0$ AFQ ordering cases, we obtain similar phase diagrams
for $V_{x^2-y^2}  = 10t$ and for $V_{3z^2-r^2} = 10t$
except for the superconducting symmetry
[Fig.~\ref{SC_AFQ_order}(a) and (b)].
We suppose that the effect of the induced $O_2^0(\bm{0})$
in the AFQ order on superconductivity is weak.
Indeed, for $V_{x^2-y^2}  = V_{3z^2-r^2} = 10t$ [Fig.~\ref{SC_AFQ_order}(c)],
the $d+id$ state without $O_2^0(\bm{0})$ realizes in a wide region.
Also for $O_2^2$ AFQ ordering cases, the effect of the induced $O_2^0(\bm{0})$
seems weak.
We obtain similar phase diagrams
for $V_{x^2-y^2}  = 10t$ and for $V_{3z^2-r^2} = 10t$
[Fig.~\ref{SC_AFQ_order}(d) and (e)],
and the $d+id$ state realizes in a wide region
for $V_{x^2-y^2}  = V_{3z^2-r^2} = 10t$ [Fig.~\ref{SC_AFQ_order}(f)].
In all of these AFQ order cases,
there are superconducting phases with and without AFQ order.

In a coexisting state of superconductivity
and staggered order with ordering vector $\bm{Q}$,
the pair amplitude with momentum $\bm{Q}$ can be finite.
In the AFQ order of $O_2^0$,
a matrix element between $(\alpha, \bm{k})$ and $(\alpha, \bm{k}+\bm{Q})$
becomes finite [see Eq.~\eqref{eq:O20}].
Here, $(\tau, \bm{k})$ denotes the electronic state
of orbital $\tau$ with momentum $\bm{k}$.
Then,
if the pair amplitude of $(\alpha, \bm{k})$ and $(\gamma, -\bm{k})$ is finite,
the pair amplitude of $(\alpha, \bm{k}+\bm{Q})$ and $(\gamma, -\bm{k})$
is also finite in the AFQ order of $O_2^0$.
In other words, the pair amplitude of $d_{x^2-y^2}$ with $\bm{Q}$ is induced
in the coexisting phase of the $d_{x^2-y^2}$ superconductivity
with $O_2^0$ AFQ order.
Similarly, the $d_{3z^2-r^2}$ superconductivity with AFQ order of $O_2^0$
accompanies the pair amplitude of $d_{3z^2-r^2}$ with $\bm{Q}$.
In the AFQ order of $O_2^2$, a matrix element between $(\alpha, \bm{k})$
and $(\beta, \bm{k}+\bm{Q})$ becomes finite [see Eq.~\eqref{eq:O22}].
Then, in the coexisting phase of the $d_{x^2-y^2}$ superconductivity
and AFQ order of $O_2^2$,
the pair amplitude of $d_{3z^2-r^2}$ with momentum $\bm{Q}$ becomes finite.
For the $d_{3z^2-r^2}$ superconductivity with AFQ order of $O_2^2$,
the pair amplitude of $d_{x^2-y^2}$ with $\bm{Q}$ is finite.
In the coexisting phase of the $d+id$ superconductivity
with $O_2^0$ or $O_2^2$ AFQ order,
the pair amplitude for $d+id$ with $\bm{Q}$ becomes finite.
We have checked these induced pair amplitudes with momentum $\bm{Q}$
in the coexisting phases
of the superconductivity with AFQ order in the numerical calculations.

\begin{figure}
  \includegraphics[width=0.99\linewidth]
      {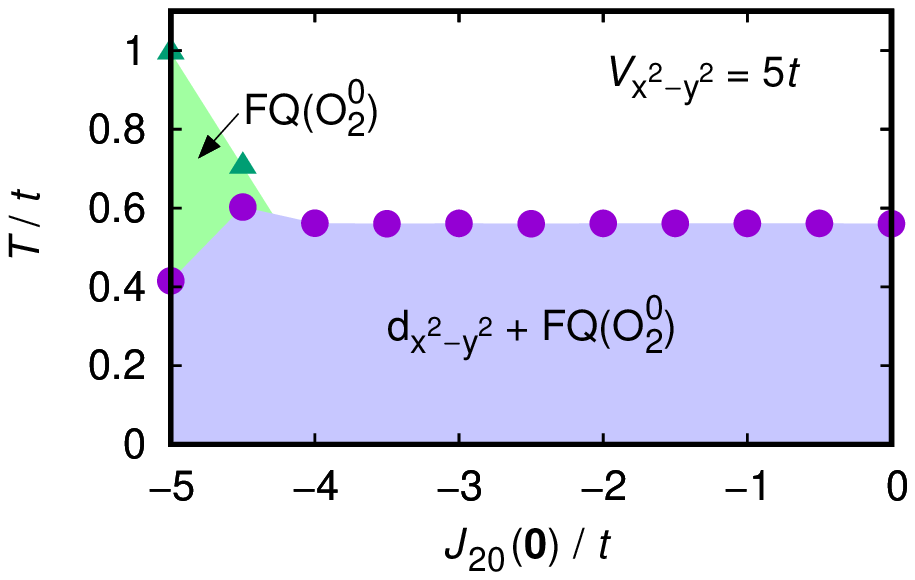}%
  \caption{
    Phase diagram of $d_{x^2-y^2}$ superconductivity
    coexisting with FQ order of $O_2^0$
    for $V_{x^2-y^2} = 5t$.
  \label{SCD13_FQO20-V13_5}}
\end{figure}

To show the dependence on the strength of the superconducting interaction,
we have constructed a phase diagram
of $d_{x^2-y^2}$ superconductivity coexisting with FQ order of $O_2^0$
for a smaller value of the superconducting interaction, $V_{x^2-y^2} = 5t$,
as an example (Fig.~\ref{SCD13_FQO20-V13_5}).
This phase diagram is qualitatively the same as for $V_{x^2-y^2} = 10t$
in Fig.~\ref{SC_FQ_order}(a).
For example, we observe an enhancement of the superconducting transition
temperature at a parameter where $T_{\text{SC}} \simeq T_{\text{FQ}}$.
However, we find that the superconductivity disappears
for a much smaller value of the superconducting interaction.
Thus, we need a certain strength of the superconducting interaction
to discuss superconductivity in the present mean-field model.

\section{Uniaxial stress}\label{stress}
To investigate the effect of uniaxial stress,
we vary the external field $H_{20}$ to the quadrupole moment $O_2^0$.
This field corresponds to a uniaxial stress along the $z$ direction.
From the definition in Eq.~\eqref{eq:H20},
positive $H_{20}$ increases $O_2^0(\bm{0})$
and negative $H_{20}$ decreases $O_2^0(\bm{0})$.

First, we consider FQ order of $O_2^0$ without superconductivity.
In this case under $H_{20} \ne 0$,
the FQ transition of $O_2^0$ is not a symmetry breaking transition
since $O_2^0$ is already finite by $H_{20}$.
Then, if the FQ transition exists, it should be of first order.
Since the FQ moment is positive without $H_{20}$,
$H_{20}<0$ suppresses $T_{\text{FQ}}$.
For $H_{20}>0$, $T_{\text{FQ}}$ can be enhanced;
however, the FQ transition becomes a crossover at a small value of $H_{20}$
since the discontinuity in the order parameter at the first order transition
is small even at $H_{20}=0$ [Fig.~\ref{quadrupole_order}(a)].

For other pure order cases,
the transition temperature increases under $H_{20}>0$ ($H_{20}<0$)
at least for a sufficiently small $|H_{20}|$
when the induced $O_2^0(\bm{0})$ is positive (negative) at $H_{20}=0$.
The transition temperature increases under $H_{20}>0$
for AFQ order of $O_2^0$
and for the superconductivity of $d_{x^2-y^2}$.
The transition temperature increases under $H_{20}<0$
for FQ of order $O_2^2$, for AFQ order of $O_2^2$,
and for the superconductivity of $d_{3z^2-r^2}$.
These behaviors seem rather trivial
and we do not show phase diagrams for these cases here.

\begin{figure}
  \includegraphics[width=0.99\linewidth]
      {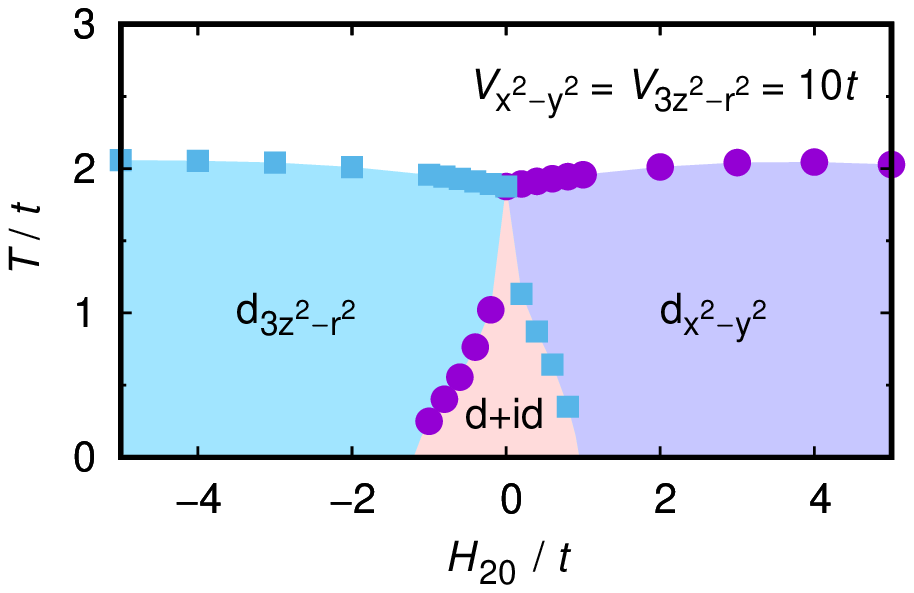}%
  \caption{
    Phase diagram of superconductivity
    in the $H_{20}$-$T$ plane
    without quadrupole interactions
    for $V_{x^2-y^2}  = V_{3z^2-r^2} = 10t$.
  \label{SCDiD_H20}}
\end{figure}

For the $d+id$ superconductivity,
$T_{\text{SC}}$ is expected to split into two by $H_{20}$~\cite{Sigrist1991}.
We demonstrate it in the present mean-field model.
Figure \ref{SCDiD_H20} shows a phase diagram in the $H_{20}$-$T$ plane
for $V_{x^2-y^2}  = V_{3z^2-r^2} = 10t$.
The transition temperature is split into two by $H_{20}$.
The higher transition temperature is slightly increased by $H_{20}$.
Below the second superconducting transition temperature,
$d+id$ superconducting state appears.
The weights of the $d_{x^2-y^2}$ and $d_{3z^2-r^2}$ components
in the $d+id$ superconducting state are different for $H_{20} \ne 0$.
For a large $|H_{20}|$, only one superconducting transition occurs.
The region of the $d+id$ superconductivity is narrow in the present model.

\begin{figure}
  \includegraphics[width=0.99\linewidth]
      {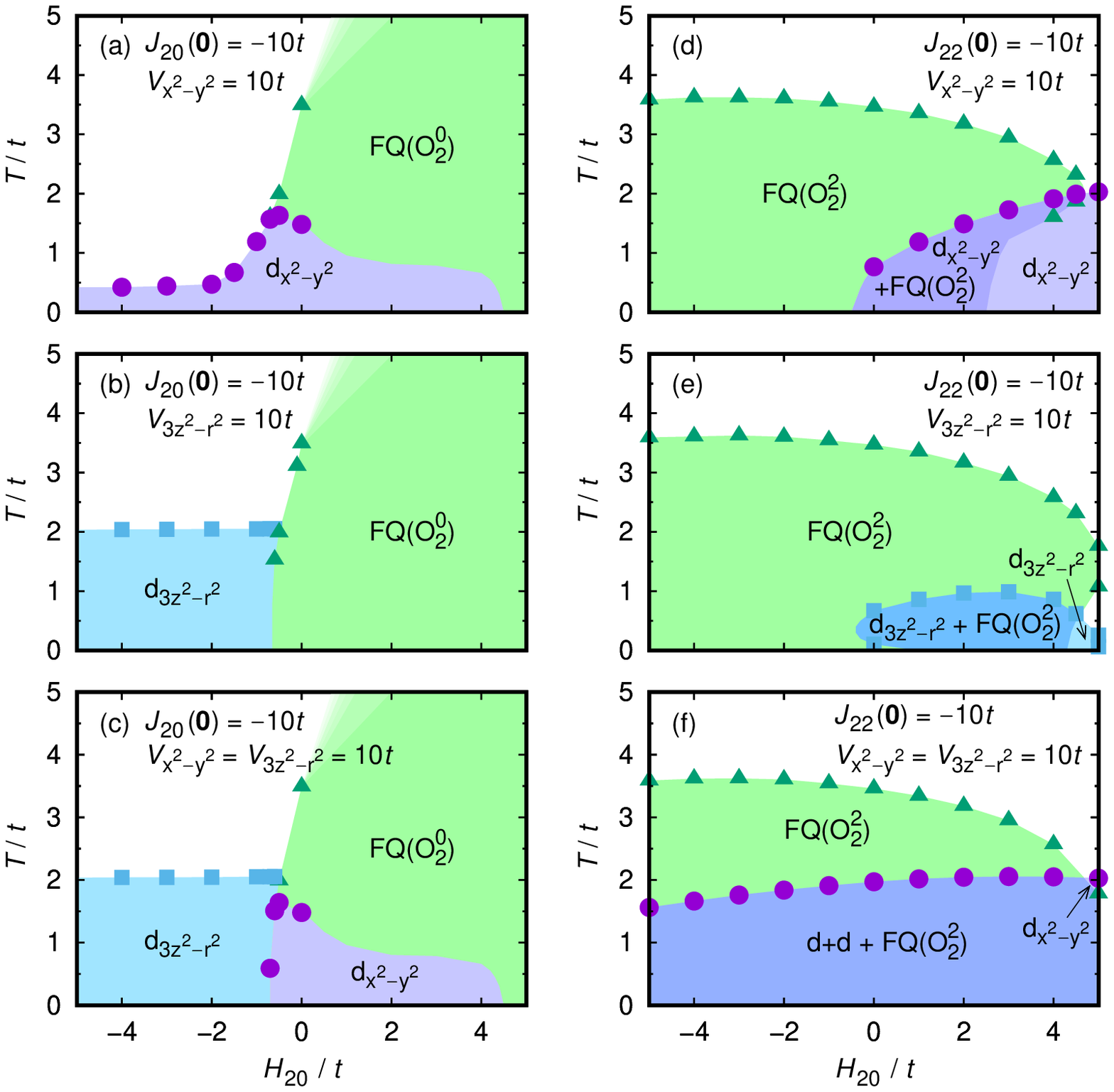}%
  \caption{
    Phase diagrams of superconductivity coexisting with FQ order
    in the $H_{20}$-$T$ plane:
    with $J_{20}(\bm{0})=-10t$
    (a) for $V_{x^2-y^2}  = 10t$,
    (b) for $V_{3z^2-r^2} = 10t$,
    and
    (c) for $V_{x^2-y^2}  = V_{3z^2-r^2} = 10t$;
    with $J_{22}(\bm{0})=-10t$
    (d) for $V_{x^2-y^2}  = 10t$,
    (e) for $V_{3z^2-r^2} = 10t$,
    and
    (f) for $V_{x^2-y^2}  = V_{3z^2-r^2} = 10t$.
  \label{SC_FQ_order_H20}}
\end{figure}

Next, we discuss the effects of the uniaxial stress
on the coexisting phases of superconductivity with FQ order.
Figure~\ref{SC_FQ_order_H20}(a) is a phase diagram
for $J_{20}(\bm{0})=-10t$ and $V_{x^2-y^2}=10t$.
For $H_{20} \gtrsim 0$, the FQ transition of $O_2^0$ becomes a crossover.
At low temperatures for $H_{20} \gtrsim 0$,
we find a solution with the $d_{x^2-y^2}$ superconductivity
but it does not fulfill the condition $n_7=n_8=1$.
It implies that phase separation occurs.
While the $d_{x^2-y^2}$ superconductivity may survive to some extent
for $H_{20}>0$,
we should consider possible phase separation for this region.
For $H_{20}<0$,
the FQ order of $O_2^0$ is suppressed and $T_{\text{SC}}$ initially increases.
By decreasing $H_{20}$ further, both the FQ order of $O_2^0$
and $d_{x^2-y^2}$ superconductivity are suppressed
as for the pure ordered phases of them.
For $J_{20}(\bm{0})=-10t$ and $V_{3z^2-r^2}=10t$ [Fig.~\ref{SC_FQ_order_H20}(b)],
we find only the FQ phase of $O_2^0$ for $H_{20}>0$
since a positive $H_{20}$ suppresses the $d_{3z^2-r^2}$ superconductivity.
On the other hand, for $H_{20}<0$, the FQ order of $O_2^0$ is suppressed
and the $d_{3z^2-r^2}$ superconductivity is realized in a wide region.
For $J_{20}(\bm{0})=-10t$ with $V_{x^2-y^2}=V_{3z^2-r^2}=10t$
[Fig.~\ref{SC_FQ_order_H20}(c)],
we obtain a phase diagram like a combination of
Fig.~\ref{SC_FQ_order_H20}(a) and (b).
Then, we expect a change of the superconducting symmetry
by applying uniaxial stress for this case.

Figure~\ref{SC_FQ_order_H20}(d) is a phase diagram
for $J_{22}(\bm{0})=-10t$ and $V_{x^2-y^2}=10t$.
For $H_{20}>0$, the FQ order of $O_2^2$ is suppressed
and $d_{x^2-y^2}$ superconductivity develops.
There are two $d_{x^2-y^2}$ superconducting phases
with and without FQ order of $O_2^2$.
For $J_{22}(\bm{0})=-10t$ and $V_{3z^2-r^2}=10t$ [Fig.~\ref{SC_FQ_order_H20}(e)],
the FQ order is suppressed and $d_{3z^2-r^2}$ superconductivity takes place
for $H_{20}>0$.
However the $d_{3z^2-r^2}$ superconductivity is also suppressed by $H_{20}>0$
and the region of this superconductivity is narrow.
Figure~\ref{SC_FQ_order_H20}(f) shows a phase diagram
for $J_{22}(\bm{0})=-10t$ with $V_{x^2-y^2}=V_{3z^2-r^2}=10t$.
In the FQ ordered phase of $O_2^2$, the system is orthorhombic
and the $d_{x^2-y^2}$ and $d_{3z^2-r^2}$ superconductivities can mix.
Then, by combining these pairing states,
a $d+d$ superconducting phase extends in a wide region.

\begin{figure}
  \includegraphics[width=0.99\linewidth]
      {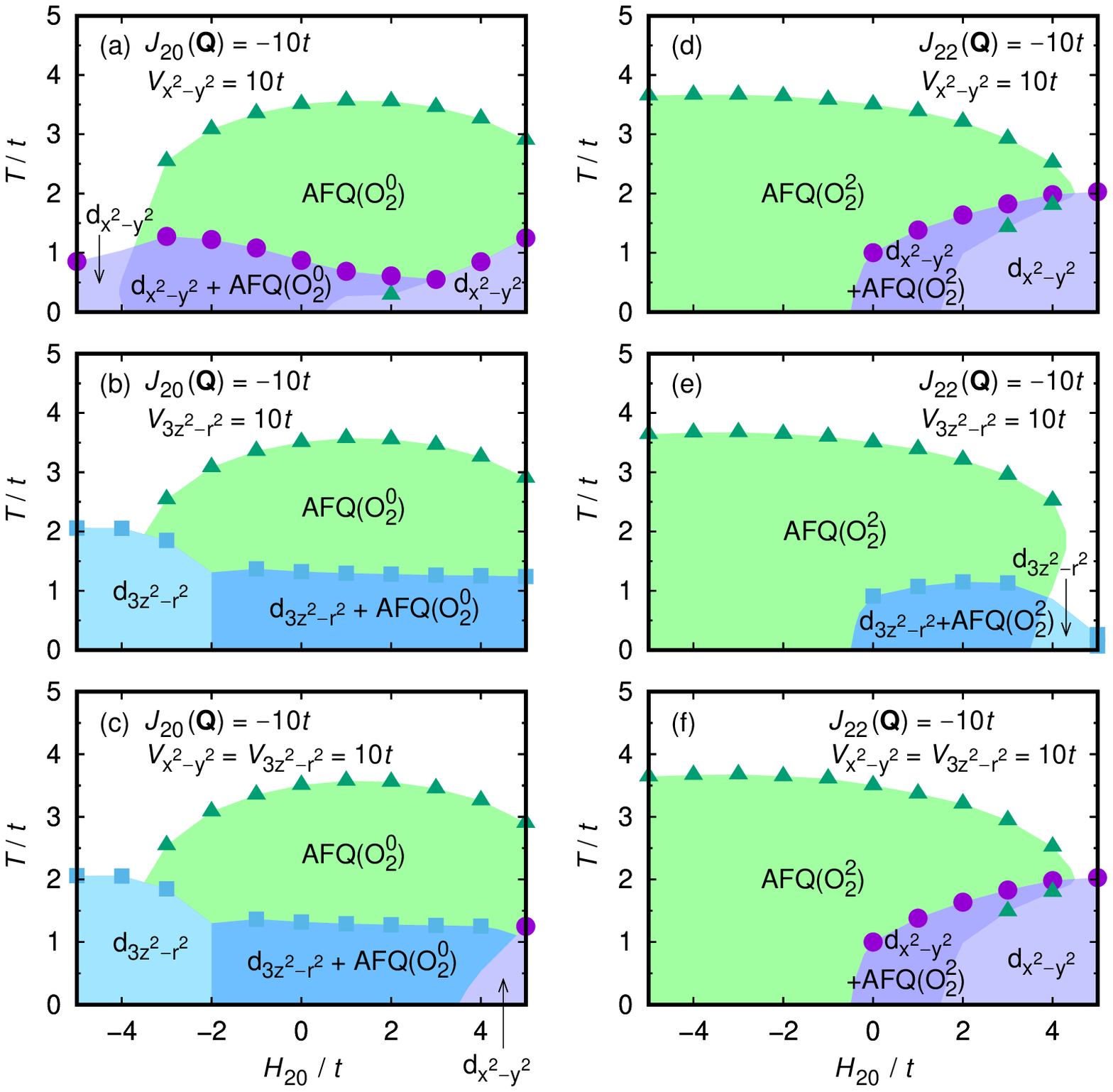}%
  \caption{
    Phase diagrams of superconductivity coexisting with AFQ order
    in the $H_{20}$-$T$ plane:
    with $J_{20}(\bm{Q})=-10t$
    (a) for $V_{x^2-y^2}  = 10t$,
    (b) for $V_{3z^2-r^2} = 10t$,
    and
    (c) for $V_{x^2-y^2}  = V_{3z^2-r^2} = 10t$;
    with $J_{22}(\bm{Q})=-10t$
    (d) for $V_{x^2-y^2}  = 10t$,
    (e) for $V_{3z^2-r^2} = 10t$,
    and
    (f) for $V_{x^2-y^2}  = V_{3z^2-r^2} = 10t$.
  \label{SC_AFQ_order_H20}}
\end{figure}

Finally, we discuss the effects of the uniaxial stress
on the coexisting phases of superconductivity with AFQ order.
Figure~\ref{SC_AFQ_order_H20}(a) is a phase diagram
for $J_{20}(\bm{Q})=-10t$ and $V_{x^2-y^2}=10t$.
For $H_{20}<0$,
the AFQ order of $O_2^0$ is suppressed and $T_{\text{SC}}$ increases.
However, a negative $H_{20}$ also suppresses
the $d_{x^2-y^2}$ superconductivity and $T_{\text{SC}}$ does not become high.
For $J_{20}(\bm{Q})=-10t$ and $V_{3z^2-r^2}=10t$ [Fig.~\ref{SC_AFQ_order_H20}(b)],
for $H_{20}<0$, the AFQ order of $O_2^0$ is suppressed
and the $d_{3z^2-r^2}$ superconductivity is enhanced.
For $J_{20}(\bm{Q})=-10t$ with $V_{x^2-y^2}=V_{3z^2-r^2}=10t$
[Fig.~\ref{SC_AFQ_order_H20}(c)],
almost the whole superconducting region is the $d_{3z^2-r^2}$ phase
since it is stable for $H_{20}<0$,
where the AFQ order of $O_2^0$ becomes unstable.
For a large $H_{20}$ the $d_{x^2-y^2}$ superconducting phase appears
since this superconductivity is cooperative with a positive $O_2^0(\bm{0})$.

Figure~\ref{SC_AFQ_order_H20}(d) is a phase diagram
for $J_{22}(\bm{Q})=-10t$ and $V_{x^2-y^2}=10t$.
For $H_{20}>0$, the AFQ order of $O_2^2$ is suppressed
and the $d_{x^2-y^2}$ superconductivity develops.
For $J_{22}(\bm{Q})=-10t$ and $V_{3z^2-r^2}=10t$ [Fig.~\ref{SC_AFQ_order_H20}(e)],
the AFQ order is suppressed and $d_{3z^2-r^2}$ superconductivity takes place
for $H_{20}>0$.
However, the $d_{3z^2-r^2}$ superconductivity is also suppressed by $H_{20}>0$
and $T_{\text{SC}}$ remains low.
Then, for $J_{22}(\bm{Q})=-10t$ with $V_{x^2-y^2}=V_{3z^2-r^2}=10t$
[Fig.~\ref{SC_AFQ_order_H20}(f)],
we obtain the same phase diagram as in Fig.~\ref{SC_AFQ_order_H20}(d)
within numerical errors.

\section{Summary}\label{summary}
We have investigated the coexisting phases of
the $d$-wave superconductivity and quadrupole order
in a model for the $\Gamma_3$ system.

In the $d$-wave superconducting states,
the cubic symmetry is broken by the anisotropic pairing
and FQ moment of $O_2^0$ becomes finite
except for the $d+id$ superconducting state.
Thus, superconductivity can occur cooperatively
with the FQ order of $O_2^0$.
Indeed, the $d_{x^2-y^2}$ superconductivity is enhanced
with the help of the FQ order in the present model.
It is in sharp contrast to ordinary cases,
where superconductivity is suppressed
when another order develops.
This case may correspond to the superconductivity
in PrTi$_2$Al$_{20}$~\cite{Koseki2011,Sakai2011,Sato2012,Ito2011,Sakai2012,
  Matsubayashi2012,Matsubayashi2014,Taniguchi2016}.

The other types of quadrupole order
are unfavorable for superconductivity.
Thus, superconductivity emerges
by suppressing the quadrupole order by reducing the quadrupole interaction
or by applying uniaxial stress.
We could not find a simultaneous transition of
superconductivity and AFQ order without fine tuning of parameters.
Thus, we need further studies to explain such a transition
observed in PrRh$_2$Zn$_{20}$~\cite{Onimaru2012}.

The $d+id$ superconductivity does not accompany a quadrupole moment
since the cubic symmetry is retained.
By lowering symmetry with a uniaxial stress,
either of the $d_{x^2-y^2}$ and $d_{3z^2-r^2}$
superconducting transition temperatures rises and at a lower temperature,
second superconducting transition breaking time reversal symmetry occurs.
Also for the other superconducting states,
the superconducting transition temperature
and superconducting symmetry can be changed by uniaxial stress.

In any case,
the effects of the hydrostatic and uniaxial pressure will be interesting
and useful to investigate superconductivity in the $\Gamma_3$ systems.
Theoretically, a microscopic description of the coexisting states
without introducing phenomenological quadrupole interactions
is also desirable.
It is an important future problem.


%

\end{document}